\documentclass{PoS}

\title{Event-by-event transverse momentum fluctuations in nuclear 
collisions at CERN SPS.}

\ShortTitle{E-by-e $p_T$ fluctuations in nucl. collisions at CERN SPS.}

\author{\speaker{Katarzyna Grebieszkow} for the NA49 Collaboration \\

        Warsaw University of Technology\\
        E-mail: \email{kperl@if.pw.edu.pl}}

\author{The NA49 Collaboration: \\
C.~Alt$^{9}$, T.~Anticic$^{23}$, B.~Baatar$^{8}$,D.~Barna$^{4}$,
J.~Bartke$^{6}$, L.~Betev$^{10}$, H.~Bia{\l}\-kowska$^{20}$,
C.~Blume$^{9}$,  B.~Boimska$^{20}$, M.~Botje$^{1}$
% ...

}

\author{ \\
$^{1}$NIKHEF, Amsterdam, Netherlands. \\
% ...
$^{4}$KFKI Research Institute for Particle and Nuclear Physics, Budapest, Hungary.\\
% ...
$^{6}$Henryk Niewodniczanski Institute of Nuclear Physics, Polish Academy of Sciences, Cracow, Poland.\\
% ...
$^{8}$Joint Institute for Nuclear Research, Dubna, Russia.\\
$^{9}$Fachbereich Physik der Universit\"{a}t, Frankfurt, Germany.\\
$^{10}$CERN, Geneva, Switzerland.\\
% ...
$^{20}$Institute for Nuclear Studies, Warsaw, Poland.\\
% ...
$^{23}$Rudjer Boskovic Institute, Zagreb, Croatia.\\
% ...

}

\abstract{The latest NA49 results on event-by-event transverse momentum 
fluctuations are presented for central Pb+Pb interactions over the 
whole SPS energy range (20$A$ - 158$A$ GeV). Two different methods are 
applied: evaluating the $\Phi_{p_{T}}$ fluctuation measure and 
studying two-particle transverse momentum correlations. The obtained 
results are compared to predictions of the UrQMD model. The 
results on the energy dependence are compared to the NA49 data on the 
system size dependence. The NA61 (SHINE, NA49-future) strategy of 
searching of the QCD critical end-point is also discussed.
}

\FullConference{Critical Point and Onset of Deconfinement
          4th International Workshop\\
		 July 9-13 2007\\
		 GSI Darmstadt,Germany}

\begin{document}

\section{Motivation}

One of the most important reasons to investigate ultra-relativistic heavy 
ion collisions is to produce and understand the properties of 
quark-gluon plasma (QGP) - a state of matter, with subhadronic degrees of    
freedom, that is expected to appear when the system is sufficiently 
hot and dense. The theoretical predictions within the Statistical Model 
of the Early Stage suggested that the energy threshold for 
deconfinement is localized 
between AGS and top SPS energies \cite{mg_model}. Indeed, the latest 
NA49 results \cite{kpi} on dependencies of various quantities on the 
collision energy seem to confirm that the onset of deconfinement sets 
in at lower SPS energies.

The phase diagram of strongly interacting matter is commonly 
presented as a ($T, \mu_B$) plot, where $T$ is the temperature and $\mu_B$ is 
a baryochemical potential. For large values of $\mu_B$ one expects a 
first order phase transition between hadron gas and QGP, which 
terminates in a critical point, and for smaller values of $\mu_B$ turns 
into a so-called crossover. According to the recent lattice QCD 
calculations, the end-point of the first-order phase transition is a 
critical point of the second-order and should be located at a 
baryochemical potential characteristic of the CERN SPS energy 
range~\cite{fodor_latt_2004}.

Dynamical (non-statistical) fluctuations are very important observables 
in the study of the phase diagram. In this proceedings article, 
transverse momentum dynamical fluctuations obtained on the basis of 
event-by-event methods will be presented. {\bf {A focus will be put on 
the energy dependence of $p_T$ fluctuations over the whole SPS energy 
range}}. The two most important reasons of studying the energy 
dependence of event-by-event $p_T$ fluctuations are:

\begin{enumerate}
\item When the studied data sample consists of dynamically very similar 
events, event-by-event fluctuations are expected to be small. In
contrast, when different classes of events are present, the fluctuations 
from one event to another are obviously much higher (various classes of 
events may exhibit different global characteristics). The latter 
situation is more probable for energies close to the phase transition 
region because QGP may be created only in a fraction of the volume of 
strongly interacting matter and this fraction can vary from event to 
event. Therefore, the energy dependence of event-by-event $p_T$  
fluctuations might exhibit enlarged fluctuations at lower SPS energies, 
where the onset of deconfinement probably occurs \cite{mg_model, kpi}.

\item Significant transverse momentum and multiplicity fluctuations 
were predicted to appear for systems that hadronize and freeze-out near 
the second-order critical QCD end-point \cite{SRS}. The phase diagram 
can be scanned by 
varying both the energy and the system size and therefore 
a non-monotonic dependence of $p_T$ and $N$ fluctuations on control 
parameters such as energy or centrality (ion size) may provide 
evidence for the QCD critical point.

\end{enumerate}

\section{Measures of transverse momentum fluctuations}

There are several methods that can be used to determine $p_T$
fluctuations on event-by-event basis. In the NA49 experiment the 
$\Phi_{p_{T}}$ fluctuation/correlation \footnote{Several effects may 
lead to non-zero value of $\Phi_{p_{T}}$. Among them are those which 
occur on an event-by-event basis (event-by-event fluctuations of the 
inverse slope parameter, existence of different event classes i.e. 
'plasma' and 'normal' events), but also 
inter-particle correlations due to Bose-Einstein statistics, Coulomb effects, 
resonance decays, flow, jet production etc.} measure, proposed in 
\cite{Gaz92}, is 
studied (for a complete definition of $\Phi_{p_{T}}$ see \cite{Gaz92} 
and the publication of NA49 \cite{fluct_size}). $\Phi_{p_{T}}$ 
quantifies a difference between event-by-event fluctuations of
transverse momentum in data and the corresponding fluctuations in 
'mixed' events. There are two important 
properties of the $\Phi_{p_{T}}$ measure. When the system consists of 
independently emitted particles (no inter-particle correlations) 
$\Phi_{p_{T}}$ assumes a value of zero. On the other hand, if A+A 
collisions can be treated as an incoherent superposition of independent 
N+N interactions (superposition model), then $\Phi_{p_{T}}$ has a 
constant value, the same for A+A and N+N interactions. 

Although $\Phi_{p_{T}}$ measures the magnitude of fluctuations it 
does not provide information on the source of underlying correlations. 
Therefore a more differential method (suggested in \cite{Tra00}) is 
also applied and two-particle correlation plots $(x_{1},x_{2})$, using
the cumulant $p_T$ variable $x$, are prepared (technical details can be 
found in \cite{fluct_size}).
Those two-dimensional plots are {\it uniformly} populated when no
inter-particle correlations are present in the system and a possible
non-uniform structure signals the presence of dynamical
fluctuations (for example, Bose-Einstein correlations lead to a ridge 
along the diagonal of the $(x_{1},x_{2})$ plot, which starts at $(0,0)$ 
and ends at $(1,1)$, whereas event-by-event temperature fluctuations 
produce a saddle shaped structure \cite{Tra00, fluct_size}).

\section{NA49, data selection and analysis}

The NA49 fixed target experiment is a large hadron spectrometer at the  
CERN SPS. The main devices of the detector are four large volume 
Time Projection Chambers (TPCs). The Vertex TPCs 
(VTPC-1 and VTPC-2) are located in the magnetic field of two 
super-conducting dipole magnets. Two other TPCs (MTPC-L and MTPC-R) 
are positioned downstream of the magnets symmetrically to the beam 
line. The NA49 TPCs allow precise measurements of particle momenta $p$ with a 
resolution of $\sigma(p)/p^2 \cong (0.3-7)\cdot10^{-4}$ (GeV/c)$^{-1}$. 
Precise measurement of specific energy loss ($dE/dx$) in the region of 
relativistic rise is possible in the TPCs, however, $dE/dx$ information 
is not used in this analysis. The centrality of the nuclear collisions 
is selected by use of information from a downstream calorimeter (VCAL), 
which measures the energy of the projectile spectator nucleons. Details 
of the NA49 detector set-up and performance of the tracking software 
are described in \cite{na49_nim}.

The data used for the analysis consists of samples of Pb+Pb collisions 
at 20$A$, 30$A$, 40$A$, 80$A$ and 158$A$ GeV energy ($\sqrt{s_{NN}}$ = 
6.27, 7.62, 8.73, 12.3 and 17.3 GeV, respectively). The fraction of the 
total inelastic cross section of nucleus+nucleus collisions 
($\sigma/\sigma_{tot}$) was set to 7.2\%. The fluctuation analysis 
presented in these proceedings is performed by use of all
charged particles, registered by the NA49 detector at forward rapidity.
Additionally, the results are prepared for negatively and 
positively charged particles, separately. In the analysis tracks with 
$0.005 < p_T < 1.5$ GeV/c are used. For all five energies the forward 
rapidity region is selected as $1.1 < y^*_{\pi} < 2.6$, where 
$y^*_{\pi}$ is the particle rapidity calculated in the center-of-mass 
reference system. As the track-by-track identification is not always 
possible in the experiment, the rapidities are calculated assuming  
pion mass for all particles.

\begin{figure}
\begin{center}
\includegraphics[width=15cm]{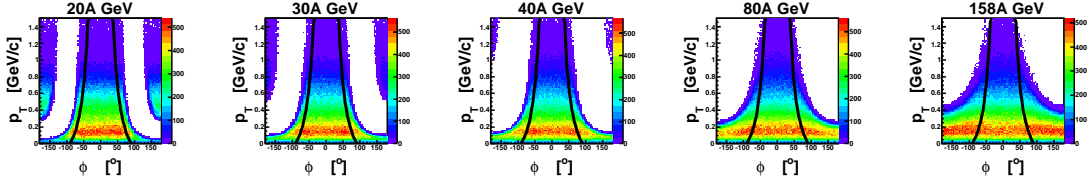}
\end{center}
\vspace{-0.8cm}
\caption {NA49 ($\phi,p_{T}$) acceptance of all charged
particles for $2.0 < y_{\pi}^{*} < 2.2$. Additional cut on
$y_{p}^{*}$ (see the text) not included. The solid lines represent the
analytical parametrization of the common acceptance.}
\label{azimuth2}
\end{figure}

Fig. \ref{azimuth2} presents examples of azimuthal angle versus 
$p_T$ ($\phi,p_T$) acceptance for all charged particles \footnote{In 
the NA49 detector positively charged particles are concentrated around 
$\phi=0 ^{o}$, whereas negatively charged ones close to $\phi= \pm 180 
^{o}$ (standard configuration of the magnetic field). Therefore in the 
plot, azimuthal angle of negatively charged particles is reflected: 
namely for particles with $\phi<0^{o}$ the azimuthal 
angle is changed as follows:$\phi=\phi+360^{o}$, and finally 
$\phi=\phi-180^{o}$.} at $2.0 < y^*_{\pi} < 2.2$. The regions of 
complete azimuthal acceptance {\it common} for all energies are 
denoted by black solid lines and described by an analytical formula:
$p_{T}(\phi)=\frac{A}{\phi^{2}}-B$, where the parameters $A$ and $B$ 
depend on the rapidity range as given in Table \ref{a_energy}. 
Only particles within the analytical curves are used in the analysis.

\begin{table}
\begin{center}
\begin{tabular}{|c|c|c|c|c|c|c|c|c|}
\hline
$y_{\pi}^{*}$  & 1.0-1.2 & 1.2-1.4 & 1.4-1.6 & 1.6-1.8 & 1.8-2.0 & 
2.0-2.2
& 2.2-2.4 & 2.4-2.6 \cr
\hline
\hline
$A [\frac{deg.^{2}GeV}{c}]$ & 600 & 700 & 1000 & 2600 & 3000 & 2500 & 
1800
& 1000 \cr
\hline
$B [\frac {GeV}{c}]$ & 0.2 & 0.2 & 0.2 & 0.5 & 0.4 & 0.3 & 0.3 & 0.1 \cr
\hline
\end{tabular}
\end{center}
\vspace{-0.5cm}
\caption {The parametrization of the NA49 $\phi - p_T$ acceptance
common for all five energies.}
\label{a_energy}
\end{table}

\begin{figure}
\begin{center}
\includegraphics[width=8cm]{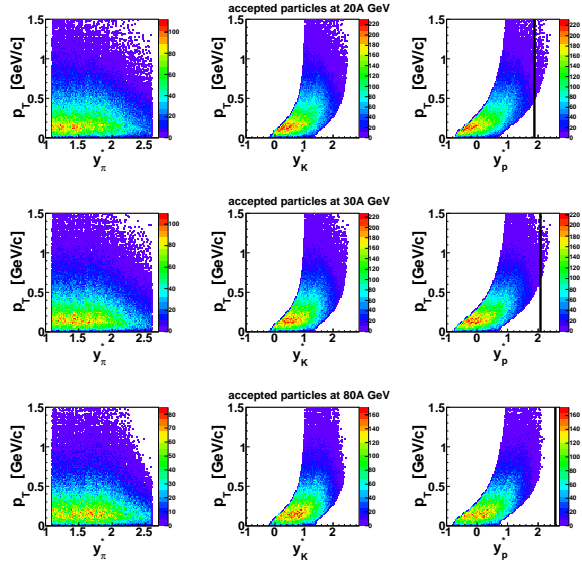}
\end{center}
\vspace{-0.8cm}
\caption{($y^{*},p_{T}$) plots of all accepted particles
assuming pion (left), kaon (middle) and proton
(right) mass. Additional cut on $y_{p}^{*}$ (see the text) is not
included. Top, middle and bottom panels correspond to 20$A$, 30$A$ and
80$A$ GeV data, respectively. Black lines represent beam rapidities
($y^{*}_{beam}$) in the center-of-mass reference system.}
\label{acceptance_ypt}
\end{figure}

Fig. \ref{acceptance_ypt} presents ($y^{*},p_{T}$) plots of all charged
particles accepted in the analysis (additional cut on $y_{p}^{*}$ -
see below - not included in the plots). It can be seen that at lower 
energies the NA49 TPC acceptance extends to the projectile spectator 
domain. This domain was excluded by an additional cut $y^*_p < y^{*}_{beam} - 
0.5$ (see below).

The methods of determining statistical and systematic errors can be 
found in \cite{fluct_size}. Systematic errors have been determined from 
$\Phi_{p_{T}}$ stability for different event and track selection criteria. The 
influence of random losses of particles (reconstruction inefficiency, 
track cuts) on $\Phi_{p_{T}}$ values has been checked and found to be 
very small in the studied kinematic region. The influence of the limited 
two-track resolution (TTR) of the NA49 detector has been determined on the 
basis of 'mixed' events and Geant simulations (details of the procedure 
can be found in \cite{fluct_size}). In the studied kinematic and 
acceptance region the values of those corrections are not higher than 4 
MeV/c (for top SPS energy) and those additive corrections 
have been applied to 'raw' $\Phi_{p_{T}}$ values.

\begin{figure}
\begin{center}
\includegraphics[width=12cm]{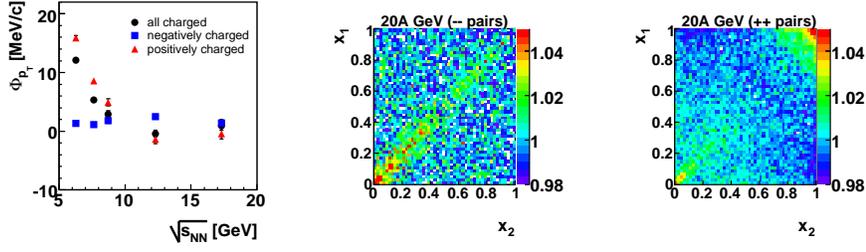}
\end{center}
\vspace{-0.8cm}
\caption{NA49 results without additional cut: $y^{*}_p <
y^{*}_{beam} - 0.5$. $\Phi_{p_{T}}$ versus energy (left) and
two-particle correlation plots $(x_{1},x_{2})$ using
the cumulant $p_T$ variable $x$ for 20$A$ GeV interactions for pairs of
negatively charged particles (middle) and positively charged ones
(right).}
\label{fipt_2D_noycut_cpod}
\end{figure}

\begin{figure}[h]
\begin{center}
\includegraphics[width=12cm]{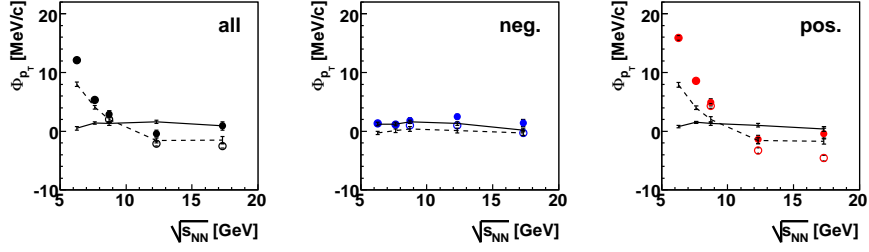}
\end{center}
\vspace{-0.8cm}
\caption {$\Phi_{p_{T}}$ as a function of energy calculated  
for {\it not} identified particles without two-track resolution 
corrections (open points) and with two-track resolution corrections 
(full points) compared to identified pions (solid curves) and 
(anti-)protons (dashed curves). Results for identified particles {\it do 
not} include two-track resolution corrections. Additional cuts on 
$dE/dx$ values and the total momentum ($p \geq 3$ GeV/c) were applied 
for proton and pion identification. The panels represent: all charged, 
negatively charged, positively charged particles, respectively. }
\label{fipt_energy_identif_p_cut}
\end{figure}

\begin{figure}
\begin{center}
\includegraphics[width=12cm]{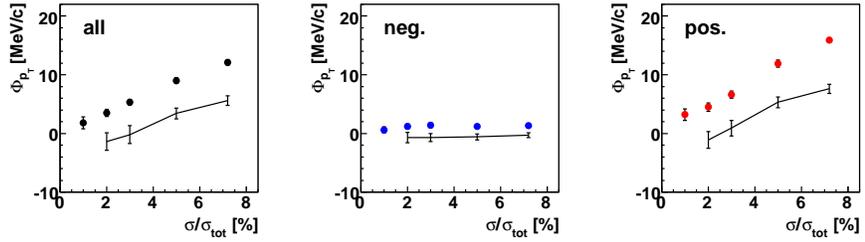}
\end{center}
\vspace{-0.8cm}
\caption {$\Phi_{p_{T}}$ for the most central 20$A$ GeV Pb+Pb 
interactions as a function of the fraction of the total inelastic cross 
section of nucleus+nucleus collisions ($\sigma/\sigma_{tot}$). Points 
represent NA49 data with kinematic and acceptance cuts as described 
above (without $y_{p}^{*}$ cut). Black lines correspond to 
the UrQMD model with the acceptance restrictions the same as for data. 
Data points are {\it not} corrected for the limited two-track 
resolution. The 
panels represent: all charged, negatively charged, positively charged 
particles, respectively. Note: the values and their errors are 
correlated.}
\label{fipt_percent_inel}
\end{figure}

\begin{figure}
\begin{center}
\includegraphics[width=12cm]{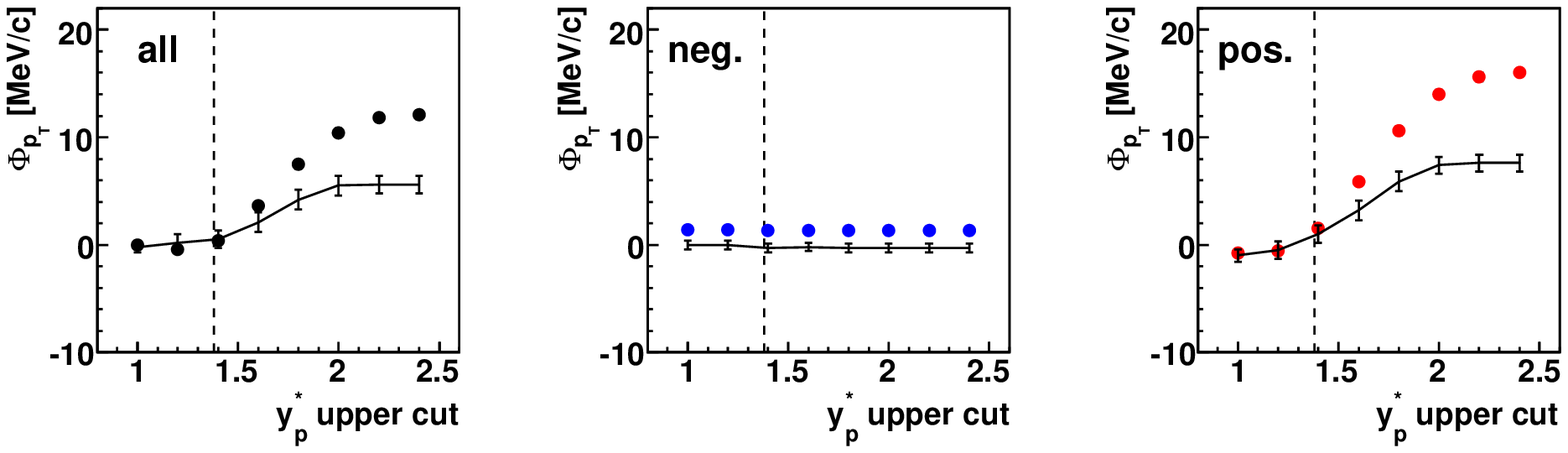}
\end{center}
\vspace{-0.8cm}
\caption {$\Phi_{p_{T}}$ as a function of an upper $y^{*}_{p}$ cut, 
obtained for 20$A$ GeV interactions. Points represent NA49 data with
kinematic and acceptance cuts as described above (without $y_{p}^{*}$ 
cut). Black lines correspond to the UrQMD model with the acceptance 
restrictions the same as for data. Data points are {\it not} corrected 
for the 
limited two-track resolution. The panels represent: all charged, 
negatively charged, positively charged particles, respectively. For 
20$A$ GeV interactions $y^{*}_{beam}$=1.88. Note: the 
values and their errors are correlated. The dashed lines indicate the 
$y_{p}^{*}$ cuts used in the analysis of 20$A$ GeV data.}
\label{fipt_yproton_cut}
\end{figure}

The preliminary analysis of the energy dependence of $p_T$ fluctuations 
showed a strong increase of $\Phi_{p_{T}}$ for {\it positively charged} 
particles at lower SPS energies (Fig. \ref{fipt_2D_noycut_cpod} (left)). 
Also two-particle correlation plots for the lowest SPS energy, exhibited 
an additional source (peak at high $x$) beyond Bose-Einstein and Coulomb 
correlations on the diagonal, but for positively charged 
particles only (Fig. \ref{fipt_2D_noycut_cpod} (right)). The UrQMD model 
qualitatively confirmed the structures observed in Fig. 
\ref{fipt_2D_noycut_cpod} (left). Moreover, both the UrQMD model and 
the NA49 data (Fig. \ref{fipt_energy_identif_p_cut}) with $dE/dx$ 
identification agree that only {\it protons} are 
responsible for the observed effect, whereas $\Phi_{p_{T}}$ for newly 
produced particles such as kaons, pions, anti-protons is consistent 
with zero. Finally, it was found that this surprising effect can be 
explained 
by event-by-event impact parameter fluctuations or more precisely by a 
correlation between the number of protons in the forward hemisphere and 
the number of protons in the production region. One can eliminate this 
trivial source of correlations either by centrality restriction (Fig. 
\ref{fipt_percent_inel}) or by rejection of the beam spectator region 
(Fig. \ref{fipt_yproton_cut}). In the analysis of the NA49 data
the rejection method is employed by applying a cut on
$y_{p}^{*}$ at each energy, i.e. the rapidity $y_{p}^{*}$
calculated with the proton mass is required to be lower
than $y^{*}_{beam}-0.5$, where $y^{*}_{beam}$ is the beam rapidity in
the center-of-mass reference system.

\section{Results and discussion, comparison to the UrQMD model}

\begin{figure}
\begin{center}
\includegraphics[width=16cm]{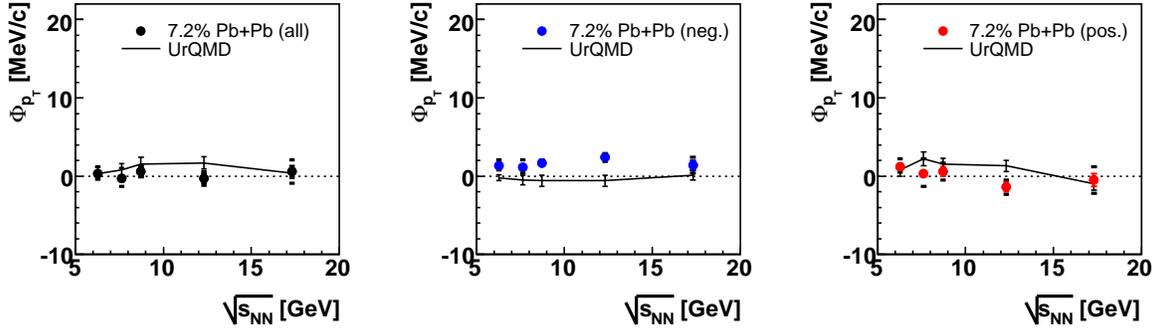}
\end{center}
\vspace{-0.8cm}
\caption {$\Phi_{p_{T}}$ as a function of energy for the 
7.2\% most central Pb+Pb interactions for all charged particles (left) 
and for negatively charged (middle) and positively charged ones (right).
Data points are corrected for the limited two-track resolution. 
Points are shown with statistical and systematic errors. NA49 results 
are compared to the UrQMD predictions (lines) with the acceptance 
restrictions.}
\label{fipt_energy1_RAP_CUT_3panels}
\end{figure}

The fluctuation measure $\Phi_{p_{T}}$, as a function of energy, is shown 
in Fig. \ref{fipt_energy1_RAP_CUT_3panels}. Three panels represent all 
charged, negatively charged and positively charged particles, 
respectively. Points correspond to data (with statistical and 
systematic errors) and lines to predictions of the UrQMD model 
\cite{urqmd1, urqmd2} with the same centrality, kinematic and 
acceptance restrictions as in the data.
For all three charge combinations no significant energy dependence of 
the $\Phi_{p_{T}}$ measure can be observed, both for data and for the UrQMD 
events. Moreover, $\Phi_{p_{T}}$ values
are consistent with the hypothesis of independent particle 
production (close to zero). The energy dependence of the $\Phi_{p_{T}}$ 
measure does not show any anomalies which might appear when approaching 
the phase boundary or the critical point.

\begin{figure}
\begin{center}
\includegraphics[width=10cm]{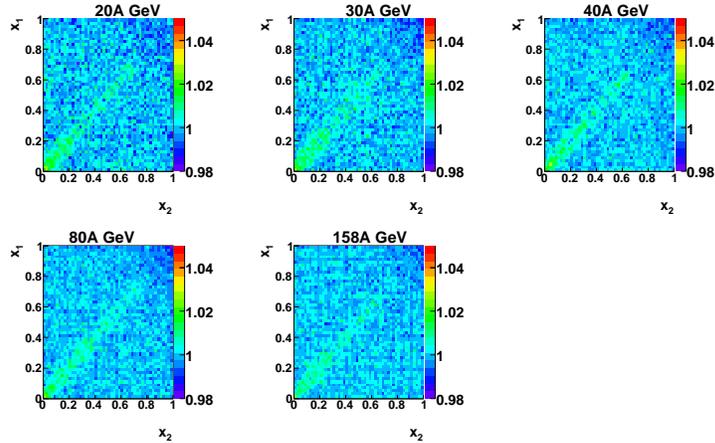}
\end{center}
\vspace{-0.8cm}
\caption {Two-particle correlation plots $(x_{1},x_{2})$ using
the cumulant $p_T$ variable $x$. The bin contents are
normalized by dividing with the average number of entries per bin.
Plots are for all charged particles produced in central Pb+Pb
collisions at 20$A$ - 158$A$ GeV.}
\label{2d_plots_energy_a_RAP_CUT}
\end{figure}

Two-particle correlation plots (for all charged particles) of the 
cumulant transverse momentum variable $x$ are presented in Fig.
\ref{2d_plots_energy_a_RAP_CUT} for 20$A$, 30$A$, 40$A$, 80$A$ and 
158$A$ GeV central Pb+Pb collisions. The plots are not uniformly 
populated but the same structure can be observed for all SPS energies. 
The enhancement of the point density in the region close to the 
diagonal is attributed to short range (Bose-Einstein and Coulomb) 
correlations.

\begin{figure}
\begin{center}
\includegraphics[width=16cm]{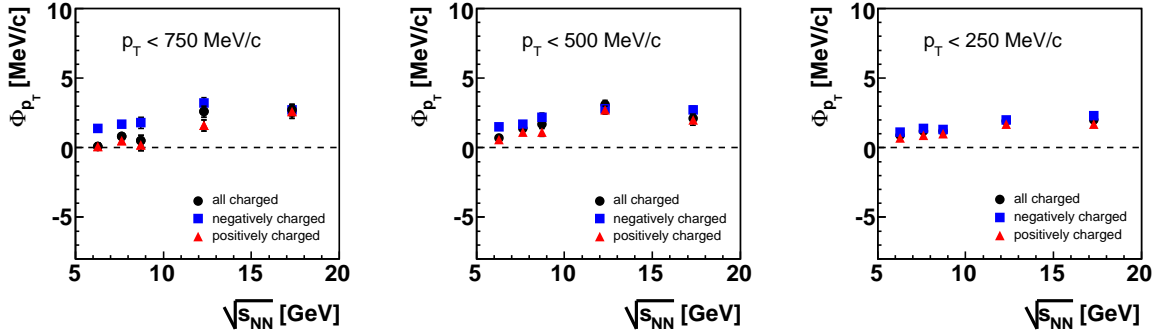}
\end{center}
\vspace{-0.8cm}
\caption {$\Phi_{p_{T}}$ as a function of energy for the
7.2\% most central Pb+Pb interactions. Results with additional cuts 
$p_T < 750$ MeV/c (left), $p_T < 500$ MeV/c (middle) and $p_T < 250$ 
MeV/c (right). Data points are corrected for the limited two-track 
resolution. Errors are statistical only.}
\label{fipt_energy_different_pt_cuts_all3}
\end{figure}

It was suggested in \cite{SRS} that fluctuations due to the 
critical QCD point should be dominated by fluctuations of pions with 
low transverse momenta (approximately below 500 MeV/c). 
Fig.~\ref{fipt_energy_different_pt_cuts_all3} shows the 
dependence of $\Phi_{p_{T}}$ on energy for several choices of an upper 
$p_T$ cut. One can see that no significant energy dependence of the 
$\Phi_{p_{T}}$ measure can 
be observed, also when low transverse momenta are selected. There are no 
anomalies exhibited and the measured $\Phi_{p_{T}}$ values are not 
significantly increased as it might be expected when the freeze-out 
takes place in the vicinity of the critical point. It should be pointed 
out that the predicted fluctuations at the critical point should result 
in $\Phi_{p_{T}}$ $\approx$ 20 MeV/c, but the effect of limited 
acceptance of NA49 (forward rapidity) reduces them to $\Phi_{p_{T}}$ 
$\approx$ 10 MeV/c \cite{SRS}.

\section{Comparison with other experiments}

\begin{figure}
\begin{center}
\includegraphics[width=6cm]{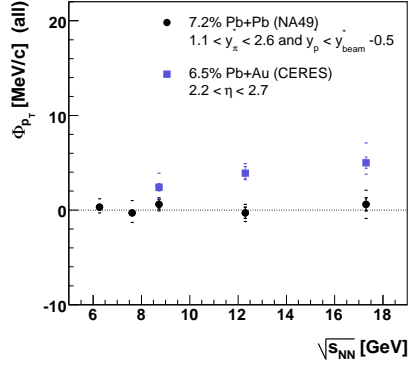}
\end{center}
\vspace{-0.8cm}
\caption {$\Phi_{p_{T}}$ as a function of energy 
measured for all charged particles by NA49 and CERES experiments. The 
NA49 points are obtained for the forward rapidity region in a limited 
azimuthal angle acceptance; the CERES data \cite{CERES} are calculated 
for the mid-rapidity region within a complete azimuthal acceptance.} 
\label{comparison_na49_ceres}
\end{figure}

Event-by-event transverse momentum fluctuations have been studied by
other experiments both at SPS and at RHIC energies. Fig. 
\ref{comparison_na49_ceres} shows the comparison of NA49 and CERES 
\cite{CERES} results on the energy dependence of the $\Phi_{p_{T}}$ 
measure. 
One can observe only very weak (if any) energy dependence of 
$\Phi_{p_{T}}$ over the whole SPS energy range. It should be however 
stressed that quantitative comparison of $\Phi_{p_{T}}$ values in NA49 
and CERES is obscured by different acceptances of both experiments 
(NA49 - forward rapidity and limited azimuthal angle, CERES - 
mid-rapidity and complete azimuthal acceptance).

Although the $\Phi_{p_{T}}$ measure seems to be close to zero at SPS 
energies, the latest STAR \cite{star_2006b} results 
show strong increase of $\Phi_{p_{T}}$ from top SPS to RHIC energies 
(up to 50 MeV/c at top RHIC energy). This effect, however,
can be related to increased contribution from (mini-)jet 
production. A non-monotonic behavior of $p_T$ fluctuations with 
collision energy (the probable effect of approaching the phase boundary 
or indication of the hadronization near the critical point) have not 
been observed neither at SPS nor at RHIC energy.

\section{Looking for the critical point in the NA61 experiment}

The latest NA49 results on the energy dependence of $p_T$ 
fluctuations in central Pb+Pb collisions seem to leave no place for 
anomalies suggestive of an 
approach to the phase boundary and for effects of the critical point.
However, the NA49 experiment measured significant non-monotonic 
evolution of $\Phi_{p_{T}}$ with the system size at top SPS energy 
\cite{fluct_size}. This tendency was also confirmed by the CERES 
experiment \cite{ceres_qm2004}. Moreover, an 
increase of multiplicity fluctuations for peripheral Pb+Pb interactions 
(when compared to p+p and central Pb+Pb collisions) was measured by NA49 
\cite{mryb}. Both observations \footnote{Simultaneous observation of 
fluctuations in $p_T$ and multiplicity with a maximum at similar 
$\langle N_W \rangle$ (mean number of wounded nucleons) 
as predicted for the critical point.} might be the first indication of 
the critical point.

\begin{figure}
\begin{center}
\includegraphics[width=10cm]{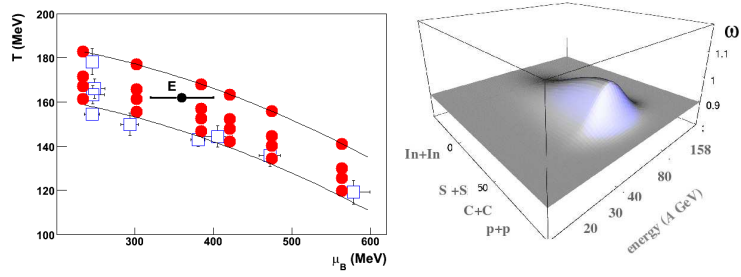}
\end{center}
\vspace{-0.8cm}
\caption {Left: Hypothetical positions of the chemical freeze-out 
points in the NA61 calculated using parametrization in \cite{becattini}. 
Predictions for InIn, SS, CC, pp (from bottom to top) and for 158A, 
80A, 40A, 30A, 20A, 10A GeV (from left to right). Open squares 
correspond to existing NA49 data. Position of the critical point (E) 
taken from \cite{fodor_latt_2004}. Right: Outline of the critical point 
search strategy in NA49 and NA61 experiments. The scaled variance 
$\omega$ \cite{mryb} represents multiplicity fluctuations.}
\label{tmiubpoints}
\end{figure}

The above results provided powerful arguments for a new experiment at 
CERN SPS - NA61 (SHINE), which is an already approved successor of the 
NA49 \cite{na49f_proposal, newSPSprog}. The NA61 experiment plans to study 
collisions of light and intermediate mass nuclei in order to cover a 
broad range of the phase diagram (Fig. \ref{tmiubpoints} (left)). Fig. 
\ref{tmiubpoints} (right) shows schematically critical point search 
strategy. Critical point can lead to an increase of $N$ and $p_T$ 
fluctuations provided the freeze-out takes place in its vicinity 
($\Delta T\approx $ 10 MeV, $\Delta \mu_B \approx $ 50 MeV 
\cite{crit_region}) and therefore the results from NA49 and NA61 may 
show a 'hill' of fluctuations over the smoothly varying background.

\begin{figure}
\begin{center}
\includegraphics[width=16cm]{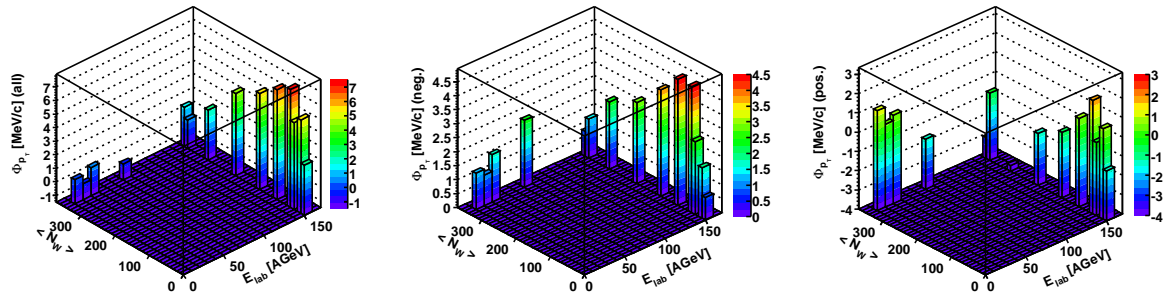}
\end{center}
\vspace{-0.8cm}
\caption {$\Phi_{p_{T}}$ as a function of energy and 
number of wounded nucleons for all charged (left), negatively charged 
(middle) and positively charged (right) particles. Errors are not shown. 
Note: different color scales; different azimuthal angle acceptance for 
the energy scan and for the system size dependence at 158$A$ GeV.}
\label{NwElab_all3panels}
\end{figure}

\begin{figure}
\begin{center}
\includegraphics[width=16cm]{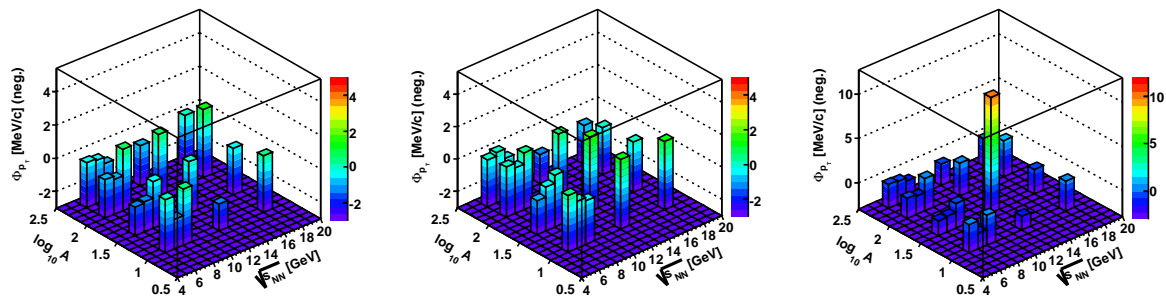}
\end{center}
\vspace{-0.8cm}
\caption {$\Phi_{p_{T}}$ as a function of energy and
atomic number (A) for forward rapidity negatively charged particles 
obtained from the UrQMD model. Statistical errors (not shown) are on 
the level of 0.5 - 1.0  MeV/c. Left: Calculations for common (for 
all energies and systems), limited azimuthal angle 
acceptance (the same as described in this paper). Middle: 
Calculations with no azimuthal angle restrictions. Right: the same 
$\Phi_{p_{T}}$ values as in the left panel, but value for 80A GeV S+S 
interactions is artificially increased by 10 MeV/c - a magnitude 
predicted by theorists for the NA49 acceptance (forward rapidity) 
\cite{SRS}.}   
\label{fipt_logAs_cpod}
\end{figure}

Fig. \ref{NwElab_all3panels} presents the current status of the NA49 
analysis, where the system size dependence at the top SPS energy is 
shown, as well as the energy dependence for 
the most central interactions over the whole SPS energy range. The 
analysis of the system size dependence at 40$A$ GeV is currently in 
progress. In Fig. \ref{fipt_logAs_cpod} the UrQMD predictions for 
the NA61 data are shown. The computation has been prepared for 
negatively charged 
particles at forward rapidities. The right panel of Fig. 
\ref{fipt_logAs_cpod} demonstrates simplified expectations for the 
presence of the critical end-point.

The latest promising results from SPS experiments motivated also RHIC 
to decrease energies down to the values $\sqrt{s_{NN}}$ = 5-15 GeV
\cite{critRHIC_plans}. Suitably to the RHIC project the JINR and GSI
laboratories intend to increase their energies that would allow for a
formation of a strongly interacting mixed quark-hadron phase
\cite{JINR_plans, FAIR_plans}. The full
review of heavy-ion facilities, of course, should mention the LHC
machine which will provide heavy-ion interactions at nearly baryon-free
region i.e. at very low $\mu_B$ values. The future results,
together with the existing data, would allow to cover a broad range in
the ($T$, $\mu_B$) plane and may help to confirm, discover or rule out
the existence of the critical point in the SPS domain.

\vspace{1cm}

{\bf Acknowledgments} \\
This work was supported by the US Department of Energy Grant 
DE-FG03-97ER41020/A000, the Bundesministerium fur Bildung und Forschung, 
Germany, 
the Virtual Institute VI-146 of Helmholtz Gemeinschaft, Germany,
the Polish State Committee for Scientific Research (1 P03B 006 30, 1 
P03B 097 29
, 1 PO3B 121 29, 1 P03B 127 30),
the Hungarian Scientific Research Foundation (T032648, T032293, 
T043514),
the Hungarian National Science Foundation, OTKA, (F034707),
the Polish-German Foundation, the Korea Science \& Engineering 
Foundation (R01-2
005-000-10334-0),
the Bulgarian National Science Fund (Ph-09/05) and the Croatian Ministry 
of Scie
nce, Education and Sport (Project 098-0982887-2878).

\end{document}